 \documentclass[final,3p,times]{elsarticle}
\usepackage{amsmath,amssymb,empheq}

\usepackage{lineno,hyperref}

\usepackage{graphics}
\usepackage{epsfig}

\usepackage[T1]{fontenc}
\usepackage[utf8]{inputenc}

\usepackage{xspace}

\usepackage{epstopdf}  % figuras eps
\usepackage{subfigure} % figuras múltiplas, com (a), (b), (c), etc.
\usepackage{wrapfig}

\usepackage{caption}
\usepackage{lscape}
\usepackage{multirow}
\usepackage{indentfirst}
\usepackage{float}
\usepackage{makeidx}
\usepackage{pifont} % simbolos legais
\usepackage[normalem]{ulem}

\usepackage[table]{xcolor}  % colorir tabelas

\usepackage{booktabs}
\usepackage{color}

\definecolor{Red}{rgb}{0.7,0.0,0.0}
\definecolor{Green}{rgb}{0.0,0.7,0.0}
\definecolor{Blue}{rgb}{0.0,0.0,0.7}

\journal{Physica A}

\begin{document}

\begin{frontmatter}



\title{Long-range correlation studies in deep earthquakes global series}


\author[ifrj]{Douglas S. R. Ferreira\corref{cor}}
\ead{douglas.ferreira@ifrj.edu.br}

\author[ifrj,ufrrj]{Jennifer Ribeiro}
%\author[ifrj]{Jennifer Ribeiro}
\ead{jennifer.conceicao@ifrj.edu.br}

\author[ifrj,uff]{Paulo S. L. Oliveira}
%\author[ifrj]{Paulo S. L. Oliveira}
\ead{paulo.oliveira@ifrj.edu.br}

\author[ifrj]{André R. Pimenta}
\ead{andre.pimenta@ifrj.edu.br}

\author[ifrj]{Renato P. Freitas}
\ead{renato.freitas@ifrj.edu.br}

\author[on,uerj]{Andr\'es R. R. Papa}
\ead{papa@on.br}

\cortext[cor]{Corresponding authors}

\address[ifrj]{LISComp Laboratory, Instituto Federal do Rio de Janeiro - Paracambi, RJ, Brazil}

\address[ufrrj]{Universidade Federal Rural do Rio de Janeiro - Seropédica, RJ, Brazil}

\address[uff]{Universidade Federal Fluminense - Niterói, RJ, Brazil}

\address[on]{Geophysics Department, Observat\'orio Nacional, Rio de Janeiro, RJ, Brazil}

\address[uerj]{Physics Department, Universidade do Estado do Rio de Janeiro, Rio de Janeiro, RJ, Brazil}

\begin{abstract}
In the present paper we have conducted studies on seismological properties using worldwide data of deep earthquakes (depth larger than 70\,km), considering events with magnitude $m \geq 4.5$.  We have  addressed the problem under the perspective of complex networks, using a time window model to build the networks for deep earthquakes, which present scale-free and small-world features. This work is an extension of a previous study using a similar approach, for shallow events. Our results for deep events corroborate with those found for the shallow ones, since the connectivity distribution for deep earthquakes also follows a $q$-exponential distribution and the scaling behavior is present. Our results were analysed using both, complex networks and Nonextensive Statistical Mechanics, contributing to strengthen the use of the time window model to construct epicenters networks. They reinforce the idea of long-range correlations between earthquakes and the criticality of the seismological system.
\end{abstract}

%%%%Graphical abstract
%\%end{graphicalabstract}

%%%%%Research highlights


\end{frontmatter}

%\linenumbers


\section{Introduction}
\label{intro}
Several phenomena in nature exhibit characteristics of complex systems such as nonlinear dynamics, fractal dimensions, power law distributions and long-range spatiotemporal memory, where the earthquakes being one example. 
The natural seismic phenomenon has the power to produce huge impacts in the human society, once that the results coming from that phenomenon can be devastating, thus the search for a better understanding of seismic dynamics becomes highly important.
One way to improve the understanding about seismological properties is the analysis of the statistical behavior of earthquakes data.
The process of seismological data mining can lead to find the arising of many patterns, which have been studied over the time  many researchers, e.g. the celebrated laws known as the Gutenberg-Richter law \cite{gutenberg1942}, for the relationship between the magnitude and the frequency of earthquakes, and the Omori law, for the rate of aftershocks following a main shock \cite{omori1894aftershocks}.

Theories of complex systems have been applied to many different areas of knowledge for a long time, such as economics \cite{foster2001competitive,hidalgo2009building}, computer science \cite{adamic2000power}, mechanical engineering \cite{pukdeboon2016anti}, biology \cite{anchang2009modeling,baianu2007categorical} and chemistry \cite{vlad2009kinetic}. 
In the last decades, works have used complex systems theories to perform studies on spatiotemporal properties of seismicity. 
In that sense, we can highlight the important approach of using complex networks concepts, which provides powerful procedures to analyse the interactions and correlations between elements of complex systems, giving efficient descriptions about the dynamics of such systems. 
The earthquake network formulation was developed by Abe and Suzuki \cite{abe2004scalefree} to define the evolving network associated with successive earthquakes. 
After that, a large number of works have used earthquake networks to study different properties of the complex seismic phenomenon, for both real  \cite{abe2004small,Abe2006,Abe2007,Abe2011,Lotfi2012,Pasten2018,Chorozoglou2019,he2019statistical} and synthetic data produced by computational models \cite{Peixoto2004a,Peixoto2004b,Peixoto2006,Caruso2006,ferreira2015agreement}.

%The second approach takes into account the study of the earthquake phenomenon from the perspective of nonextensive statistical mechanics, specially regarding the analysis of signatures found in probability distributions of spatial and temporal characteristics of earthquakes. 
%Nonextensive statistical mechanics (NESM) is a theory very useful to explain complex systems where the standard Boltzmann–Gibbs statistical mechanics does not seem to apply, i.e., systems with strong interactions between their elements, long-range spatiotemporal correlations and fractal geometries. 
%Using concepts from NESM, several researchers have investigated spatial, temporal and intensity properties of the seismicity in many different places, e.g., California \cite{AbeSuzuki2003distance,AbeSuzuki2005timeinterval,dias2019analysis}, Japan \cite{AbeSuzuki2003distance,AbeSuzuki2005timeinterval}, Iran \cite{Darooneh2008,Darooneh2010}, Greece \cite{Michas2013,Vallianatos2014}, Brazil \cite{Silva2006,Scherrer2015} and Mars \cite{Vallianatos2013}. 
%Similarly, the NESM was also used in artificial data produced by computational models for simplified earthquakes reproduction \cite{ferreira2015agreement,Caruso2007}.

However, despite the various studies made using concepts of complex systems, we have a very limited set of works considering earthquakes occurred in large geographic regions, specially for the entire world. 
As reported by different authors, many studies suggest that earthquakes may induce other earthquakes several hundred miles away from the rupture zone, since there is no consensus about a physical length scale for aftershocks zones \cite{steeples1996far,Baiesi2004,baiesi2005complex,paczuski2005spatial,Abe2012,Bendick2017}. 
Considering that perspective, Ferreira \textit{et al.} \cite{ferreira2014small} have analysed natural seismic events for the globe, investigating the \textit{epicenter network} (where two regions will be connected by an edge if two successive events occur in those respective regions) and the time interval between successive earthquakes (also referred as ``calm time'' or ``inter-event time''). 
The network resulting from this process has been found to being scale-free and small-world-like, with the probability distribution for the time interval between successive earthquakes presenting nonextensive characteristics.  It is noteworthy that Nonextensive Statistical Mechanics (NESM) is a theory very useful to explain complex systems where the standard Boltzmann–Gibbs statistical mechanics does not seem to apply, i.e., systems with strong interactions between their elements, long-range spatiotemporal correlations and fractal geometries. 
Recently, in \cite{Ferreira2018}, the authors have improved the work done in \cite{ferreira2014small} by modifying the way to build the epicenter network. 
To do that they have defined a ``time window'' to connect different regions over the world, where the region corresponding to the first event is connected to all regions within this window by directed edges but respecting the time order of earthquakes. 
Thereby, using data for shallow earthquakes for the entire world, the authors find nonextensive characteristics also in the epicenter network, keeping scale-free and small-world properties.

%Even though the last two works mentioned before indicate that the nonextensive behavior is also present when we look at the seismic events from a global perspective, those works have considered only a few spatiotemporal features.
%In that way, in the present paper we have used data from the worldwide earthquake catalog for the period between 2002 and 2016, considering separately the shallow events from the deep ones, to carry out a more complete study on the dynamics of the earthquakes using temporal and spatial probability distributions of earthquakes. 

Nevertheless, there is no such work made for deep earthquakes, which restricts the complex network analysis of the worldwide seismological phenomenon. Furthermore, it is important to highlight that, in South America, for example, intermediate-deep earthquakes produced the highest numbers of fatalities of all earthquakes in the twentieth century \cite{frohlich2006deep}, pointing once again the importance of studying deep earthquakes.

Following the methodology used in \cite{Ferreira2018} for shallow earthquakes, in the present work we will consider only deep earthquakes to build the epicenter network in order to calculate topological properties that characterize the network, producing a complementary study about epicenter network studies on earthquakes all over the world.
We have analysed our results under the perspective of complex network and NESM theories, also taking into consideration arguments of \textit{finite data-size scaling}. 

In the next sections we present the seismological data used and the methodology employed to construct the epicenter network as well as our results, analyses and considerations.
%In Section \ref{nesm}, aiming to make this paper as self-contained as possible, we present a brief theoretical background on nonextensive statistical mechanics, to introduce important theories and equations that will be used throughout this work. 

%Previous works have
%Complex networks applications.
%NESM applications.
%Limitations, i.e., what other do not have done yet.
%What we have done in previous works (especially in EPL paper).
%What we are doing at the present paper.
%The paper sections.

\section{Data Selection}
\label{data}

The seismological data used to perform our analyses was taken from the World Catalog of Earthquakes of Advanced National Seismic System (ANSS)\footnote{https://earthquake.usgs.gov/data/comcat/} covering the entire world for the period between 2002 and 2016. 
The magnitude types considered were $M_b$ (body-wave magnitude), $M_d$ (duration magnitude), $M_l$ (local magnitude) and $M_w$ (moment magnitude).
It is relevant to highlight that we have considered only earthquakes with magnitudes larger than or equal to 4.5, once in that catalog events with magnitudes less than 4.5 are not completely registered for the whole world. 
In addition, we have also disconsidered all non-earthquake events, e.g. quarry blasts, sonic booms, rock bursts and nuclear explosions. With that the total number of events in our dataset is 101\,746.

At this point it is noteworthy that we have separated shallow from deep earthquakes data. 
In the present study we will analyse only the deep ones, i.e., the events occurred at depths greater than 70\,km, once the shallow earthquakes have already been studied in \cite{Ferreira2018}.
That separation has the objective to make comparisons between earthquakes with similar seismic origins. 
After this differentiation we get a number of 21\,226 deep events. 
In order to perform a preliminary check of our data, we have verified that earthquakes' magnitude data follows the Gutember-Richter law with a \textit{b-value} exponent equal to $1.08$, as expected that value is close to $1.00$, since we have excluded events with small magnitudes.

\section{Method}
\label{method} 

As previously mentioned, the technique used in the present paper, to study our dataset, is the construction of complex networks from earthquake data.
To construct and analyse the epicenter network of our data we have followed the same method proposed in \cite{Ferreira2018}, so this section is dedicated to present a brief review and explanation of that method. 

The most commonly method used for the construction of earthquake complex networks is based on the method proposed by Abe and Suzuki \cite{abe2004scalefree}, where the geographical region under consideration is divided into cells of the same size, which will become vertices of the network if earthquake epicenters occur therein. 
Two successive events define an edge between two vertices. 

However, a new procedure to build the epicenter network was proposed by Ferreira \textit{et al.} \cite{Ferreira2018}, because the authors have observed that the standard method of construction, which considers edges only between successive events, presents problems in the cumulative probability distribution of connections for the global network, indicating a loss in the connectivity ($k$) of vertices with large number of connections \cite{ferreira2014small}. 
A summary of that procedure is as follows. First, the Earth's surface is divided into equal square cells of side $L \times L$ where a cell will become a vertex of the network every time the epicenter of an earthquake is located therein. The spatial determination of the cell of each epicenter is made by using the latitude $\theta_E$ and longitude $\phi_E$ of each epicenter relative to a given reference, $\theta_0$ and $\phi_0$. Adopting those coordinates and using the spherical approximation for the Earth (with radius $R=6.371\times 10^3$\,km), we can calculate the distances North-South ($S^{ns}_E$) and East-West ($S^{ew}_E$) by 
\begin{equation}
\label{distance}
\begin{array}{l}   % they are two tiny letters "L"  inside the keys, to indicate "left-justified"
S^{ns}_E = R.\theta_E
\\
S^{ew}_E = R.\phi_E.\cos{\theta_E},
\end{array}
\end{equation}
where, for the sake of simplicity, we have considered $(\theta_0, \phi_0) = (0,0)$. We can also note that, regardless the region of the planet, the cells have always the same size, $L \times L$.

Secondly, it is defined a \textit{time window} to create the connections (edges) between the vertices. 
Keeping the chronological order, the time window, $T$, is placed on the data so that the first vertex inside the window is connected to all other vertices within that window. 
After that, the window is moved forward to the next event and the connections procedure is replicated. 
It is repeated until it is no longer possible to move the window forward. 
It is worth mentioning that the viewpoint adopted with this construction mechanism allows us to better study the long-range relationships between the cells, since it takes into consideration more intensely the relationships between each cell inside the window, which does not happen for the network construction model of successive events.

\begin{figure}[t]
\begin{center}
\includegraphics[width=0.65\columnwidth]{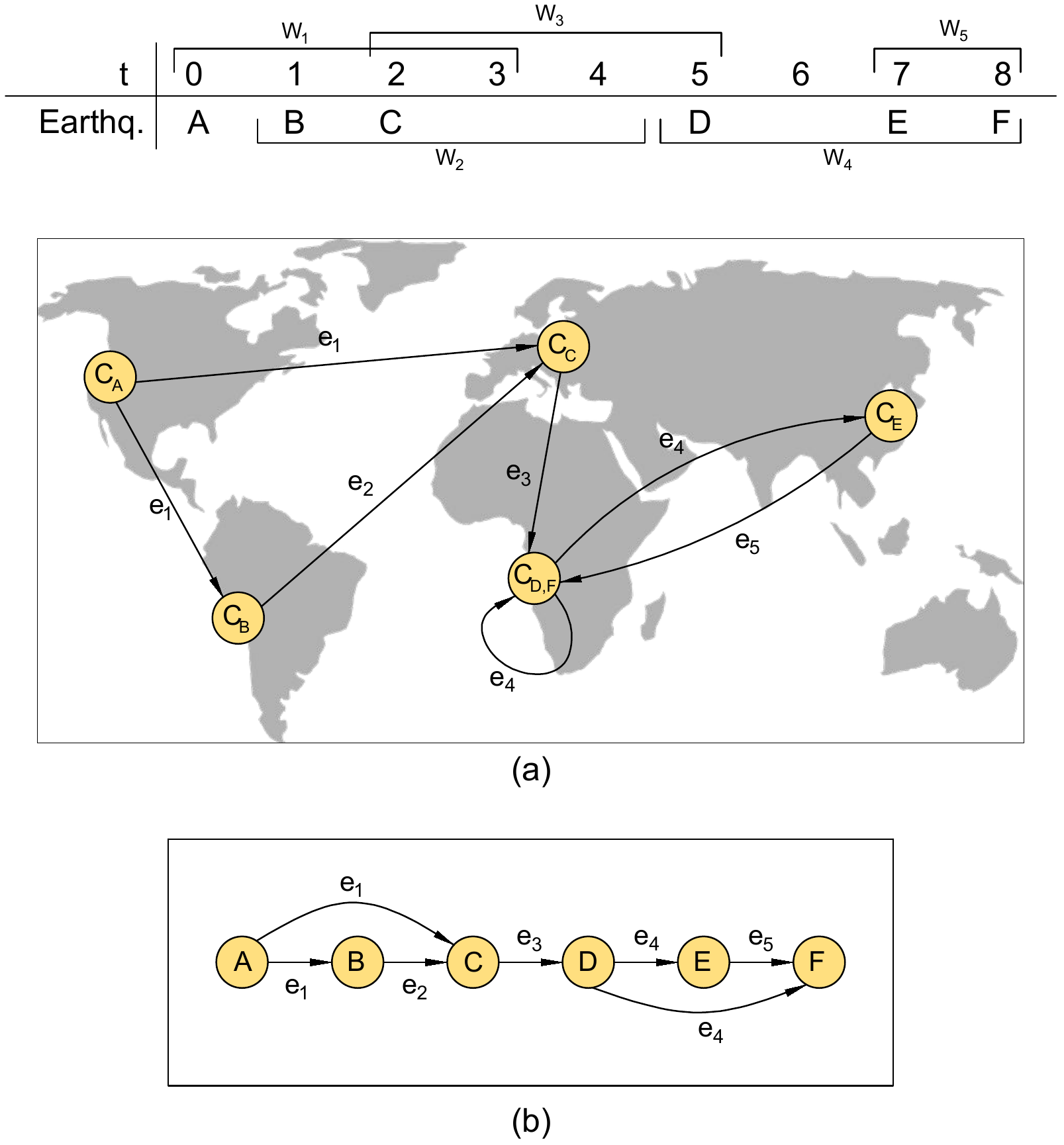}
\caption{Example of the networks construction applying a time window. The time windows are represented by $w_i$, where $i$ is the window number and all time windows have the same value, $T = 3$ (in an arbitrary unit). In (a) we have the \textit{epicenter network}, where $C_A, C_B, ..., C_F$ represent the cells that became network vertices. In (b) it is shown the corresponding \textit{earthquake network}, where the vertices are the events themselves. In both figures $e_n$ are the edges between the vertices.}
\label{fig:map}
\end{center}
\end{figure}

Thirdly and finally, it must be find the best time window size to be used in epicenter network construction. 
To accomplish that it is necessary to define an \textit{earthquake network}, which is different from the \textit{epicenter network}.
The earthquake network is a network of connections between the events and the epicenter network is a spatial network of cells where the earthquakes occurred. 
Fig~\ref{fig:map} brings to us an exemplification about the difference between \textit{epicenter network} and \textit{earthquake network}.
By applying different time window values in data to construct the respective \textit{earthquake networks} and by using the Louvain method to calculate the number of communities\cite{blondel2008fast}, it is possible to observe that the number of communities in the \textit{earthquake networks} has a maximum value regarding the time window size. 
It means that at this maximum the network has a configuration where the connections between the vertices are distributed in the better possible way in order to detect long-range correlations between the events and cells.
We highlight here that the ``best time window'' is a window not too large, which would create many links between earthquakes that probably are not correlated, but neither too small, once it would brings to us a very fragmented network in which many earthquakes that have correlations between them are not connected.

Once the best time window value is found in the \textit{earthquake network}, this value is also applied to the spatial \textit{epicenter network}, since this will be the network used for study the spatiotemporal correlations.
%Thus, the time window is an agent that acts in a way to improve the method of connections between the elements of the network, since the value of the window will define what elements will be connected.

%%%%%%%%%%%%%%%%%%%%%%%%%%%%%%%%%%%%%%%%%%%%%%%%%%%%%%%%%%%%%%%%%%%%%%%%%%%%%%%%%%%%%%%

\section{Results and discussion}
\label{results}

%\red{GUTEMBERG-RICHTER B-VALUE PARA PROFUNDO??}

%consider larger thresholds of magnitude the number of data drops drastically.

As mentioned before, a way to improve our knowledge about the earthquake dynamics is through the spatiotemporal analysis of earthquakes data. 

\begin{figure}[t]
\begin{center}
\includegraphics[width=0.55\columnwidth]{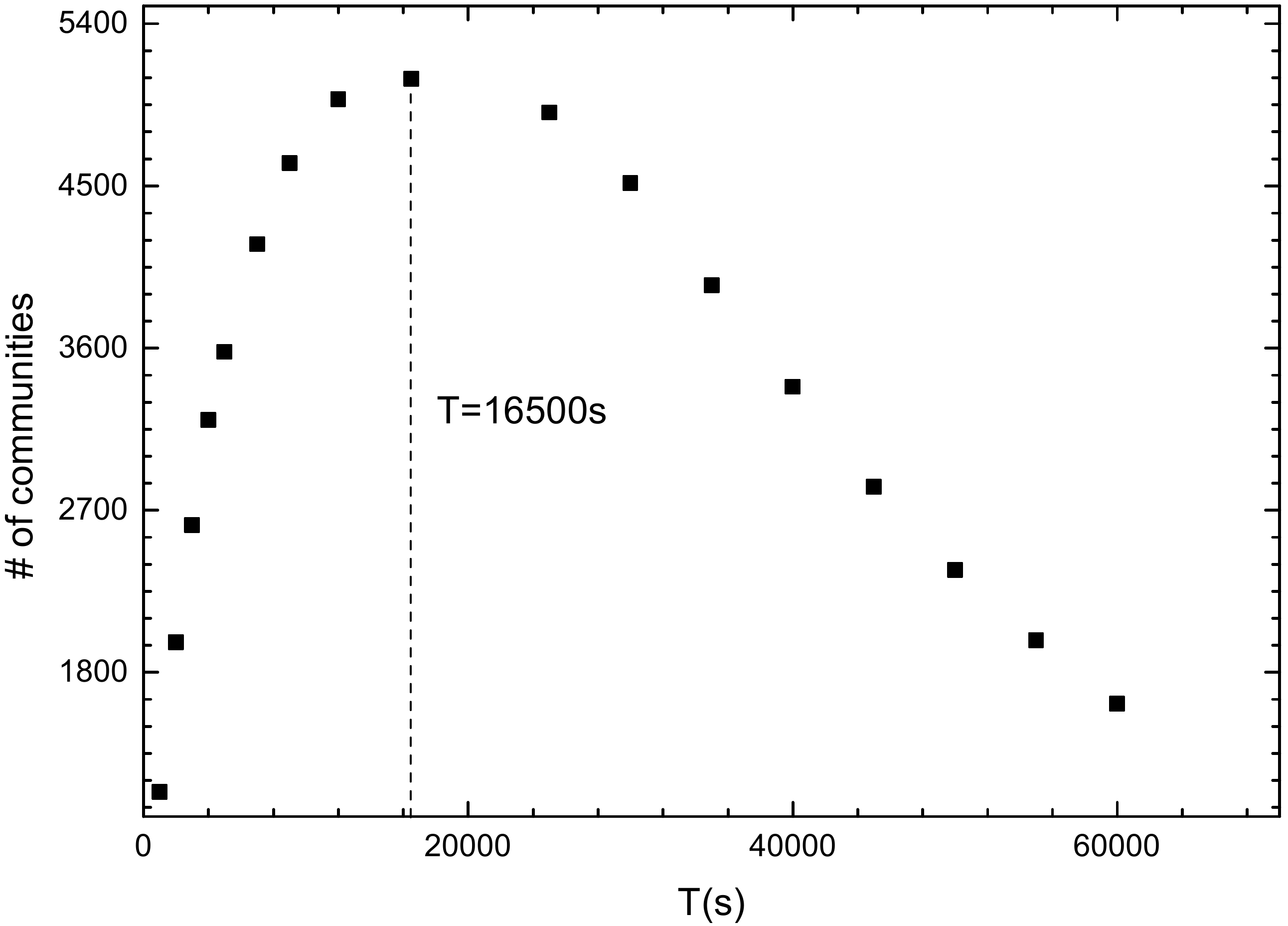}
\caption{Number of communities in the \textit{earthquake networks} for each time window size. The seismic data range from 2002 to 2016, with $m \geq 4.5$. The best time window is found for $T$\,=\,16\,500\,s, where the number of communities is maximum.}
\label{fig:communities}
\end{center}
\end{figure}

The viewpoint we have implemented to study the seismic phenomenon is the analysis of a complex network created from the earthquakes data.
In this work we conducted a similar study to the one done in \cite{Ferreira2018} but using data of deep earthquakes, i.e., for events with hypocenters deeper than 70\,km. 
We have used the procedure described in the previous section to build the networks and to find the best time window value for our deep earthquakes data.
In Fig~\ref{fig:communities}, it is observed that small windows will cause an excessive segmentation of the earthquakes network, represented by the low number of communities, and large windows will produce big clusters, which will also cause a decrease in the number of communities. 
In this way, the best time window value is given when the number of communities is maximum, i.e., when $T$\,=\,16\,500\,s.

In \cite{Ferreira2018}, Ferreira $et$ $al.$ showed, for the first time, that the distribution of connectivities for the worldwide network of shallow earthquakes constructed using the ``time window model'' has a scale-free behavior and it is invariant with respect to the window size. 
The scale-free behavior was also found in studies for specific regions of the world using the successive model of connections \cite{abe2004scalefree,abe2004small}.

\begin{figure}[t]
\begin{center}
\subfigure[]{
\includegraphics[width=0.46\columnwidth]{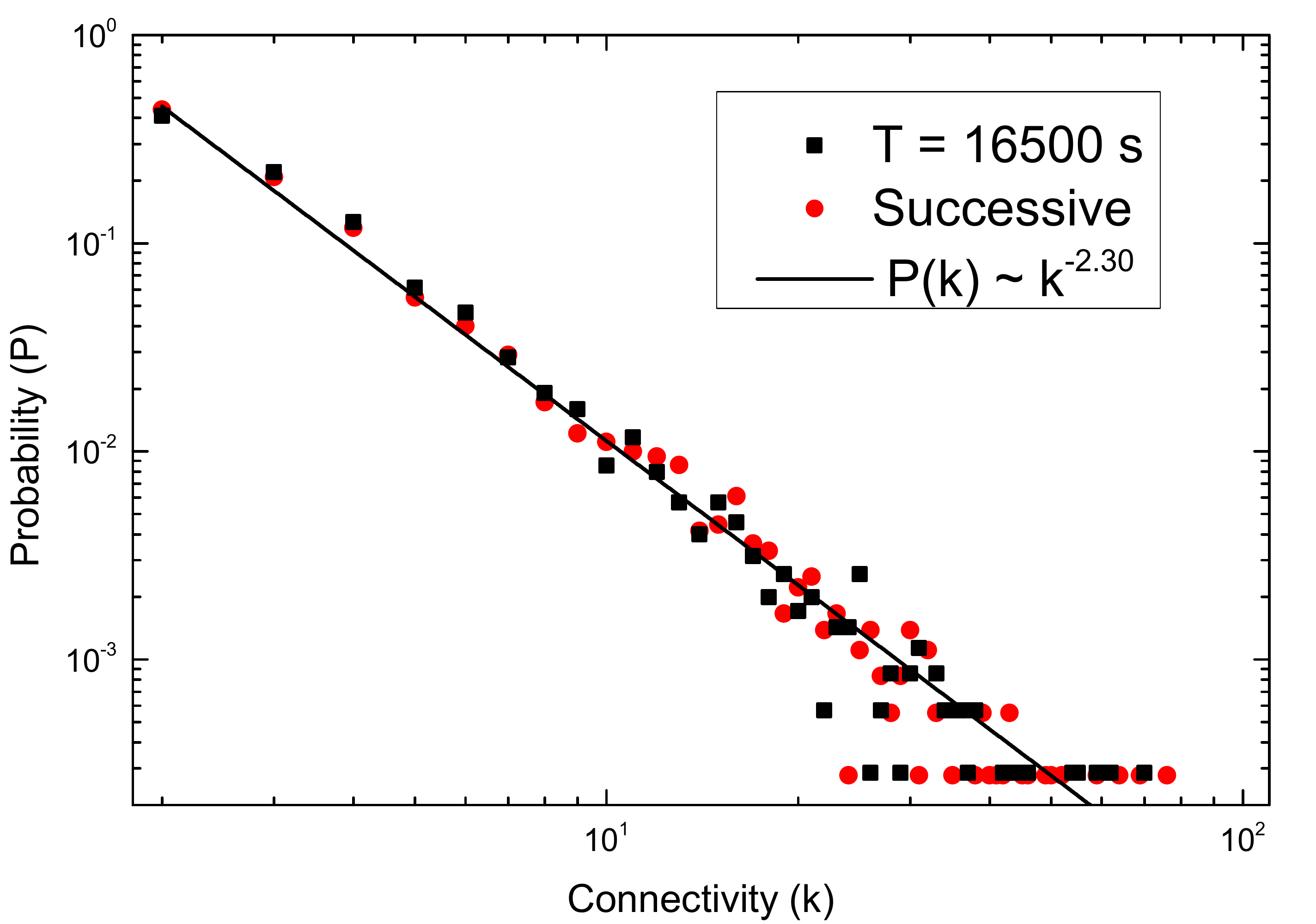}
\label{fig:k_TW_Successive}
}
\subfigure[]{
\includegraphics[width=0.46\columnwidth]{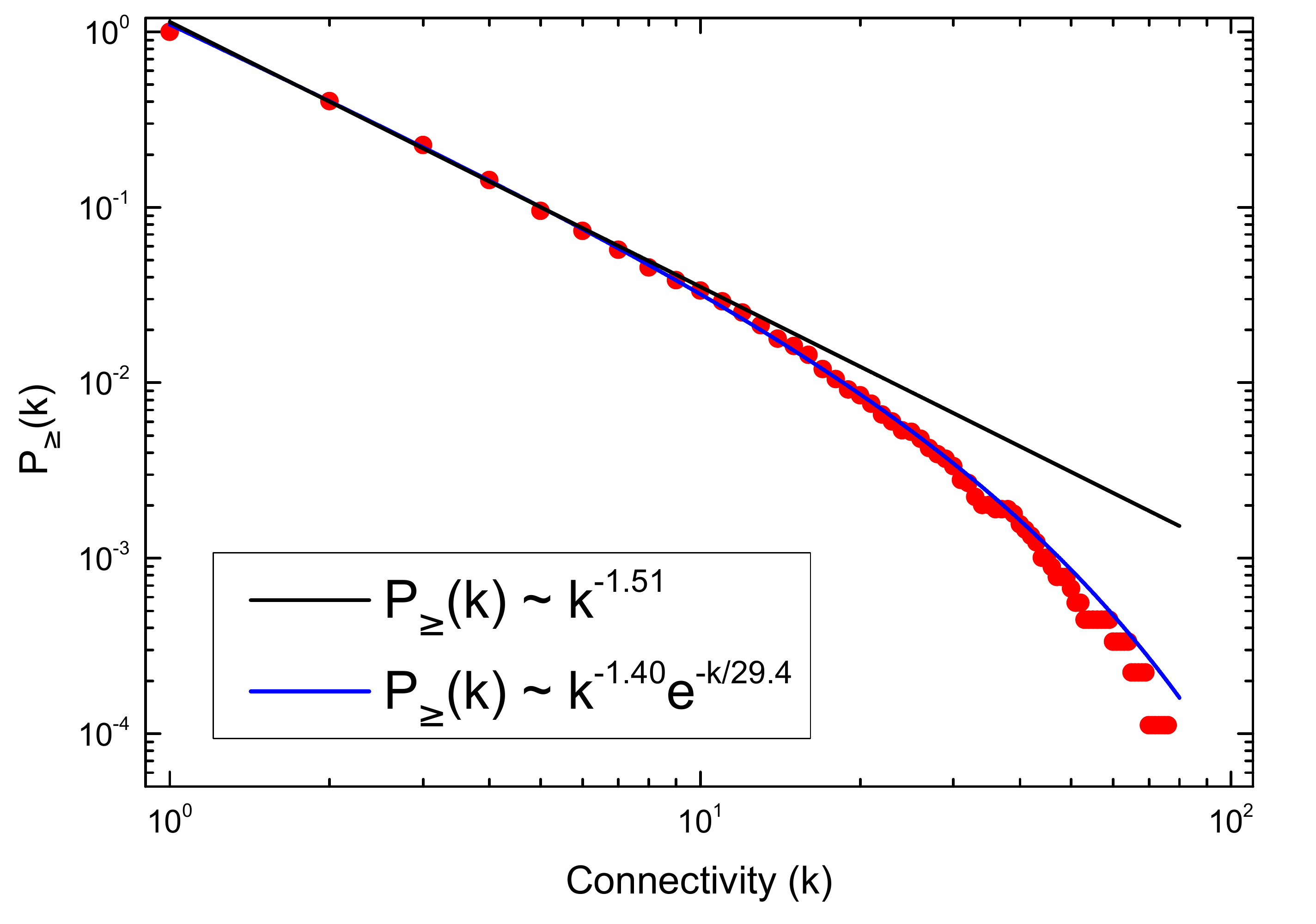}
\label{fig:k_Successive_cumulative}
}
\subfigure[]{
\includegraphics[width=0.46\columnwidth]{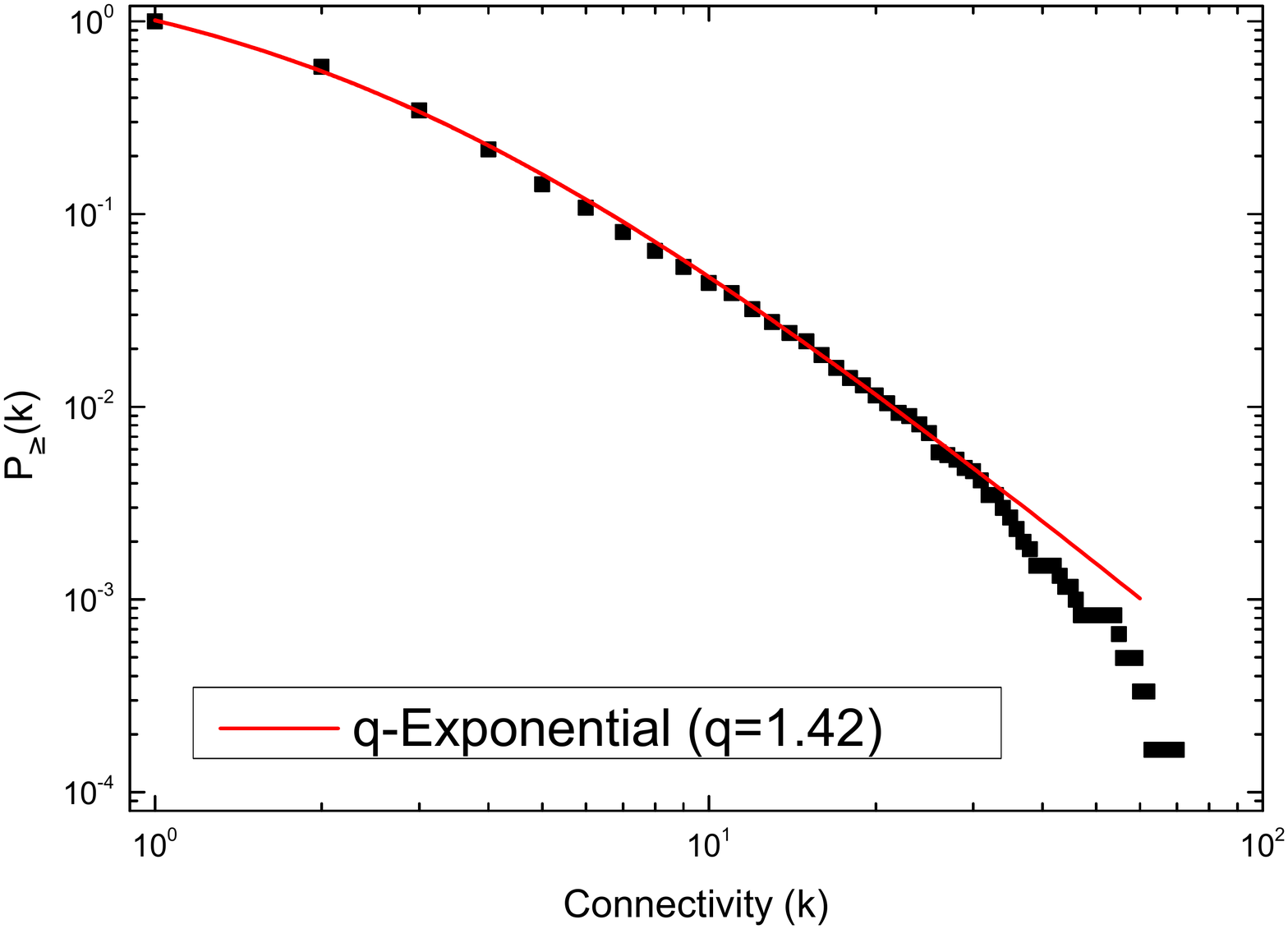}
\label{fig:k_TW_cumulative}
}
\caption{Distributions of connectivities for worldwide deep seismic data ranging from 2002 to 2016 with $L$ = 20\,km and $m \geq 4.5$.
\subref{fig:k_TW_Successive} Comparison between the time window model for $T$\,=\,16\,500\,s and the successive model. Both distributions follow a power law, $P(k) \sim k^{-\gamma}$, with $\gamma=2.30 \pm 0.02$. \subref{fig:k_Successive_cumulative} Cumulative probability distribution using the successive model. The best fit (blue line) is obtained for a power law with exponential cutoff, $P(k) \sim k^{-\alpha} e^{-k/k_c}$, with $\alpha = 1.40 \pm 0.02$ and $k_c = 29.4$. The power law regime has an exponent $\gamma=1.51$ (black line). \subref{fig:k_TW_cumulative} Cumulative probability distribution using the time window model. The best fit is obtained for a $q$-exponential function, $P_\geq(k) \sim [1 - (1 - q)\beta x]^{1/(1 - q)}$, with $\beta=0.98 \pm 0.02$ and $q=1.42$.}
\label{fig:prob_k}
\end{center}
\end{figure}

In Fig~\ref{fig:k_TW_Successive} we have the density probability distribution of connections for global networks of epicenters with cells of size 20\,km $\times$ 20\,km, constructed using both the successive and the time window models, where the last was made utilizing the best time window value found. 
In both cases it is possible to observe that the density probability distribution has a power law behavior, leading to the idea that the growth rule of this network follows a preferential attachment.

This result is not exactly the same as the one found in \cite{Ferreira2018}, since for shallow earthquakes the density probability distributions of connections for successive and time window models are different, while here, for deep events, they are very similar. 
However, as can be seen in Figs~\ref{fig:k_Successive_cumulative} and \ref{fig:k_TW_cumulative}, when we take the cumulative probability distribution we recover the same result found for shallow earthquakes, i.e., the successive and time window models have not the same behavior. A possible explanation to this fact could be that, for deep events the earthquake network (for time window model) have the majority of its cells with connectivity (\textit{in} or \textit{out}) equal to one, so this fact makes the connectivity probability distributions for time window model to be not too different from the one found for the successive model, causing that the difference between the networks built from the two different models is only noticeable when we look at the cumulative distribution.
%When using the successive model, the Fig~\ref{fig:k_Successive_cumulative} shows that the connectivity cumulative probability distribution has a pure power law behavior only for small values of $k$, because that distribution is actually better fitted by a power law with exponential cutoff, $P(k) \sim k^{-\alpha} e^{-k/k_c}$. 

%When using the successive model, the Fig~\ref{fig:k_Successive_cumulative} shows that the cumulative distribution  does not follow a pure power law, but a power law with exponential cutoff, $P(k) \sim k^{-\alpha} e^{-k/k_c}$. 

Fig~\ref{fig:k_Successive_cumulative} shows that for the successive model the connectivity cumulative distribution 
has a pure power law behavior only for small values of $k$, because that distribution is actually better fitted by a power law with exponential cutoff, $P(k) \sim k^{-\alpha} e^{-k/k_c}$. 
It is noteworthy that a very similar result was also found for California data, where the exponential cutoff does not seem to exist in probability density plots, but it appears in the cumulative ones, because in a density plot the fluctuations are higher than in a cumulative probability plot, as reported in \cite{Abe2006,ferreira2014small}.
%It is noteworthy that a very similar result was also found for California data, where the pure power law behavior appears only in the probability density plot, since the cumulative probability plot follows a power law with exponential cutoff.
%the exponential cutoff () and the q-exponential () do not seem to exist in a probability density plot, however they appear in the cumulative ones, because the fluctuations are higher than in a cumulative probability plot, as reported in \cite{ferreira2014small}.

%It is noteworthy that a very similar result was also found for California data, where the pure power law behavior appears only in the probability density plot, since the cumulative probability plot has a power law with exponential cutoff behavior \cite{Abe2006,ferreira2014small}.

When using the time window model to construct the network of epicenters, the cumulative probability plot follows the non-traditional function $q$-exponential, $P_\geq(k) = A[1 - (1 - q)\beta x]^{1/(1 - q)}$, as shown in Fig~\ref{fig:k_TW_cumulative}. This function belongs to the scope of the Nonextensive Statistical Mechanics theory 
\cite{tsallis1988possible}, which adopts the Tsallis' entropy, that can explain the statistical properties of a variety of complex systems at their quasi-equilibrium states with characteristics such as long-range interaction between its elements and long-range temporal memory. The $q$-exponential function appears naturally from the maximization of the Tsallis' entropy under appropriate constraints and it is a generalization of the exponential curve, where taking the limit $q \rightarrow 1$ we recover the standard exponential and for $q > 1$ it exhibits power law asymptotic behavior, for larger values of $k$. That result is in close agreement with the results found in previous works for shallow earthquakes \cite{Ferreira2018} and for networks built from sinthetic epicenters data produced using a small-world topology version of the OFC model \cite{ferreira2015agreement}.

\begin{figure}[t!]
\begin{center}
\subfigure[]{
\includegraphics[width=0.47\columnwidth]{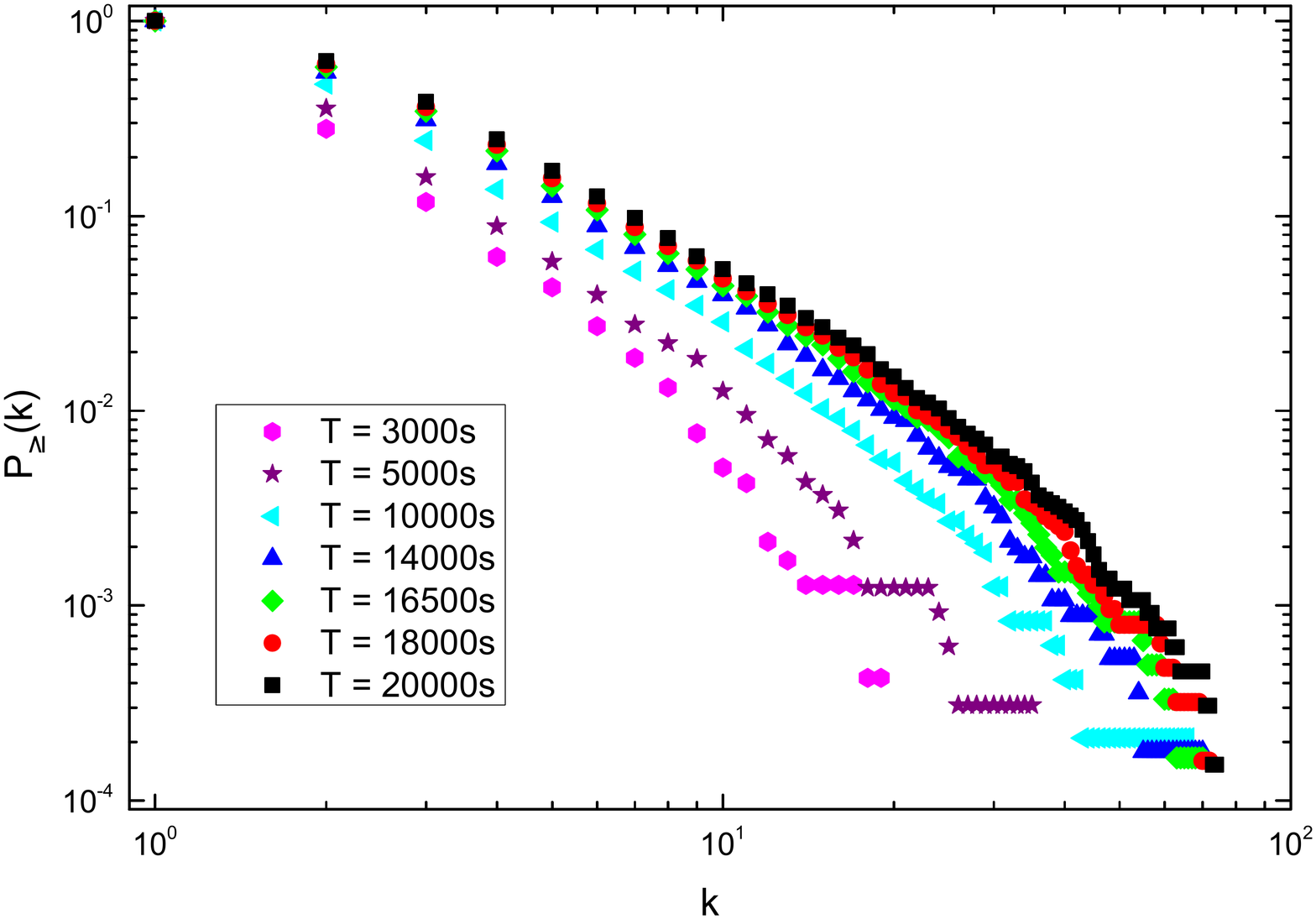}
\label{fig:k_win_TW_cumulative}
}
\subfigure[]{
\includegraphics[width=0.45\columnwidth]{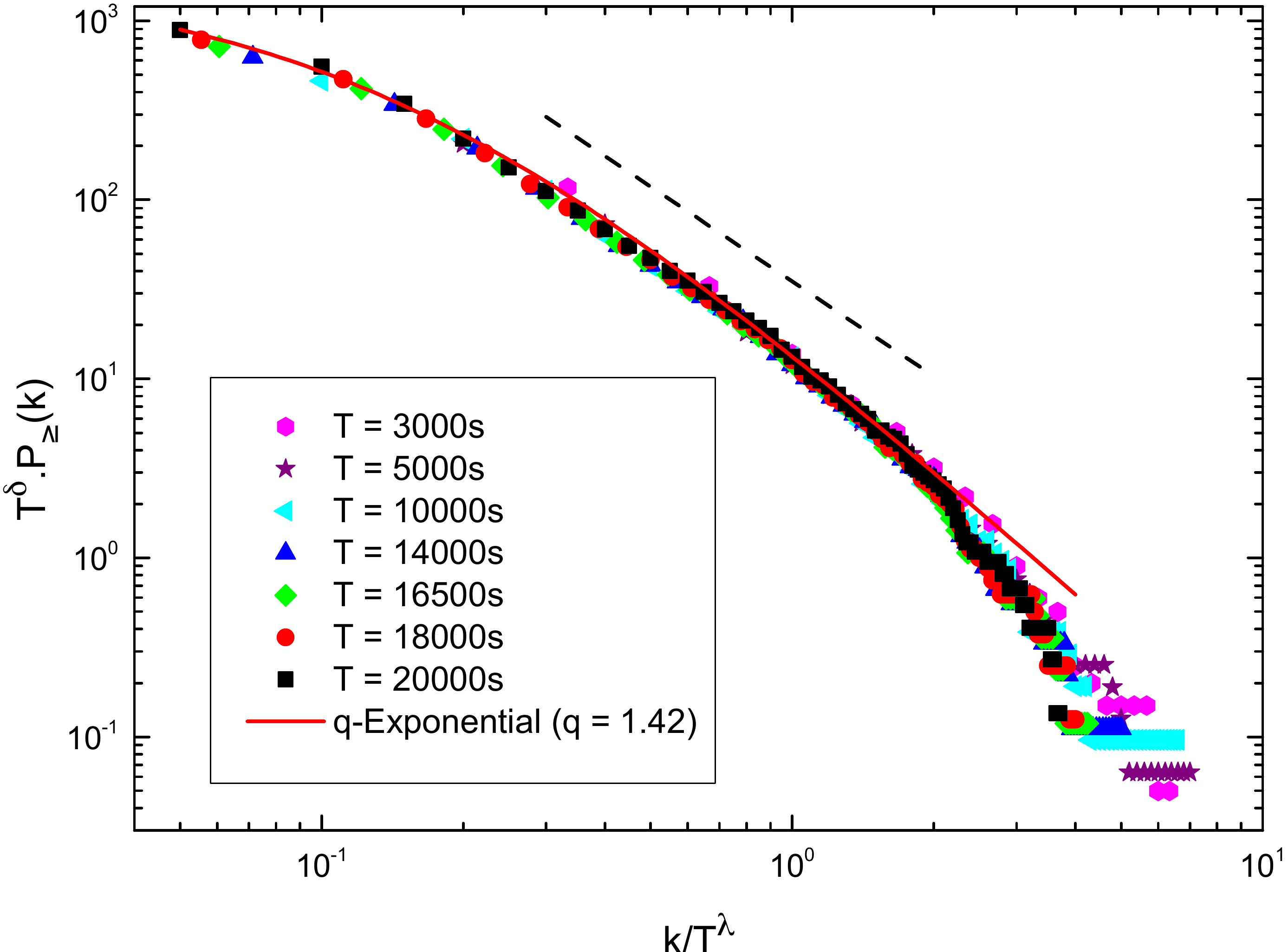}
\label{fig:k_win_TW_scaling}
}
\caption{Cumulative probability distributions of connectivities for worldwide deep seismic data ranging from 2002 to 2016 with $L$ = 20\,km and $m \geq 4.5$.
{\subref{fig:k_win_TW_cumulative} Distributions with window size ranging from $T$\,=\,3\,000\,s to $T$\,=\,20\,000\,s.
{\subref{fig:k_win_TW_scaling} The same cumulative probability distributions, but applying the scale function presented in eq. \ref{scale_function}}}. The plot presents a data collapse for many different window sizes. The best fitting is for a $q$-exponential with $\beta=16.35 \pm 0.10$ and $q=1.42$ (red line). The power law with exponent $\gamma = 1.80$ (black dashed line) is shown as a guide to demonstrate the power law  asymptotic behavior for large values of $k$.}
\label{fig:prob_k_win}
\end{center}
\end{figure}

We have also noticed that it is possible to make the connectivity distribution invariant with respect to the value of the time window by using the scale function 
\begin{equation}
\label{scale_function}
P_{\geq}(k,T) = T^{-\delta}f(k/T^{\lambda}), 
\end{equation}
where $\delta = 1.80$, $\lambda = 1.00$, $f(x)$ decays as $x^{-\gamma}$ with $\gamma = 1.80$ and the data collapse is in agreement with the scaling hypothesis, as can be seen in Fig~\ref{fig:prob_k_win}. It should be noted that for shallow earthquakes the scaling property was also present \cite{Ferreira2018}.

In order to complement our complex network study on seismological deep data we have calculated two important metrics, which are, the {\it{clustering coefficient}} ($C$) and the {\it{average shortest path}} ($\ell$). These metrics are important because they are used to characterize networks with dense connectivity areas and long jumps between these areas.
These are called \textit{small-world networks}, which have small {\it{average shortest path}} when compared to the number of vertices ($N$) and high {\it{clustering coefficient}} when compared to a similar random network.

The measurements of $C$ and $\ell$, were performed using the algorithms described in \cite{barrat2004architecture} and \cite{brandes2001faster}, respectively. 
The results for our epicenters network, using the time window model, were $C=4.91 \times 10^{-1}$ and $\ell=5.01$, with $N=7\,675$. 
For a random network with the same number of vertices the \textit{clustering coefficient} is $C_{rand}=5.88 \times 10^{-4}$. 
These results reveal that, similarly to what occurs in shallow earthquakes, deep ones also exhibit small-world characteristics in the global network of epicenters created from the time window model, given that $C \gg C_{rand} $ and that $\ell \ll N$ ($\ell \approx \ln N$) \cite{barabasi2016network}.
We point out here that the small-world property does not appear when considering the successive model for the network construction, once its \textit{clustering coefficient} is $C_{succ}=9.00 \times 10^{-3}$, while for this network $C_{rand}=5.26 \times 10^{-4}$, which means that it does not have $C_{succ} \gg C_{rand} $, despite of having a small \textit{average path length} ($\ell = 4.73$, with $N=8\,958$). 
The complex networks properties found by the time window modelling is in strong agreement with many previous works, where the small-world and scale-free features arise in networks of earthquakes \cite{abe2004small,Abe2006,Abe2007,Abe2011,Lotfi2012,ferreira2014small,Pasten2018,Ferreira2018,Chorozoglou2019,he2019statistical}.

\begin{figure}[tb]
\begin{center}
%\subfigure[]{
\fbox{\includegraphics[width=0.75\columnwidth]{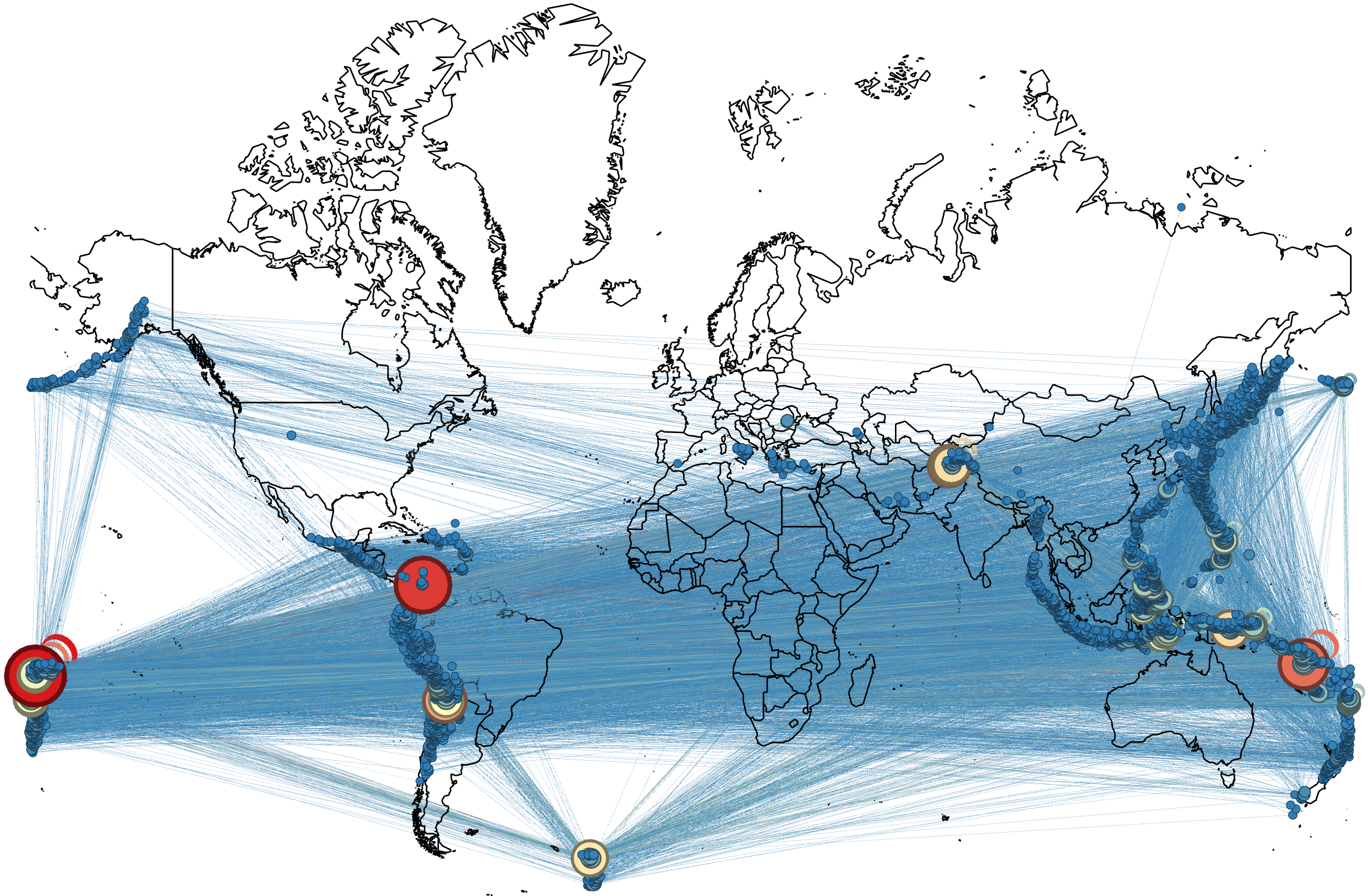}}
%\label{fig:map_tw}
%}
%\subfigure[]{
%\includegraphics[width=0.9\columnwidth]{pics/Map_successive_deep.pdf}
%\label{fig:map_suc}
%}
\caption{Geospatial picture of the world epicenter network for cells with $L$\,=\,20\,km, built from time window model for $T$\,=\,16\,500\,s. Data ranges from 2002 to 2016 and $m \geq 4.5$.  Larger and reddish cells have higher number of connections. The sites with the largest cells are located around the Fiji Islands, Colombia, Vanuatu, Argentina and Afghanistan, in that order.}
\label{fig:maps}
\end{center}
\end{figure}

Moreover, we obtained a geospatial image of our network constructed with the time window model, in order to visualize it. 
Using informations such as the exact geographical location of each cell, the number of connections of each cell and to which other cells they are connected to, it was possible to create with the software Gephi \footnote{https://gephi.org/} the image shown in Fig~\ref{fig:maps}. 
It was observed that the cells are located in regions where usually occur deep earthquakes (mostly in subduction zones at convergent plate boundaries \cite{frohlich2006deep}) and that the cells with the highest connectivity are located in the Pacific Ocean near the Fiji Islands, in Molagavita-Colombia, in the Coral Sea near Vanuatu, in Salta-Argentina and in Jurm-Afghanistan, respectively. 
It seems to correspond the geographic regions with very intense deep seismic activity, but not necessarily with the greatest earthquakes in the period we have studied. 
The biggest hub obtained in deep earthquakes, for instance, does not hold the seism with greatest magnitude in our dataset, but it is located near the Fiji Islands, belonging to the Tonga-Kermadec region, which is known for being the region where deep earthquakes occur more than in any other place around the world, being the probable cause of this remarkable activity the subduction of the 70–100\,Ma-old Pacific plate beneath the Fiji Plateau \cite{frohlich2006deep}. 
That result is different from the one found in \cite{Ferreira2018} for shallow earthquakes,  where the hubs coincide with regions that have occurred the seisms with the largest magnitudes, which makes a sense, given that deep earthquake aftershock sequences generally show aftershock productivity rates of one to three orders of magnitude less than shallow earthquakes of equivalent size, making the smaller number of aftershocks observed for deep earthquakes one of the most clear difference between shallow and deep earthquakes \cite{wiens2001seismological}.

%Moreover, we obtained a geospatial image of our network constructed with the time window model, in order to visualize it. 
%Using informations such as the exact geographical location of each cell, the number of connections of each cell and to which other cells they are connected to, it was possible to create with the software Gephi \footnote{https://gephi.org/} the image shown in Fig~\ref{fig:maps}. 
%It was observed that the cells are located in regions where it is known to often occur deep earthquakes (subduction zones at convergent plate boundaries \cite{frohlich2006deep}), showing that our result for deep earthquakes, using the time window model, makes sense. 
%\red{Fiji Island, Colombia, Argentina and Afghanistan SAO DE MAIOR MAGNITUDE?? ``our method of network construction is able to naturally identify the regions of higher seismic in- tensities using only properties of complex networks and that, through the existing connections in the network, it becomes possible to establish relations between the most diverse regions of the world'' ??}
%For comparison purposes, we have also constructed the geospatial image for the network which was built using the successive model of connections (Fig~\ref{fig:map_suc}).} 
%Comparing both images we can observe a similarity between them. 
%That similarity can be explained by...

\section{Conclusions}
\label{conclusions}

%In this work we have firstly addressed the models developed in \cite{AbeSuzuki2003distance} and \cite{AbeSuzuki2005timeinterval} to study shallow and deep worldwide earthquakes. 
%The results we have obtained with that analysis corroborate with those found by Abe and Suzuki in these previous works, where the spatial and temporal probability distributions are better fitted by $q$-exponential functions, indicating the presence of a nonextensive behavior in the spatio-temporal earthquake dynamics. 
%However our results are wider, once they are not limited to small or specific regions, but to the entire world, for both shallow and deep events.

%We have also considered the concept of ``propagation'' defined by Davidsen and Paczuski  \cite{paczuski2005spatial}, which can be represented by the introduction of a quantity similar to velocity. 
%Once again we have found results for the world in agreement with those for California, since the probability distributions of velocities, for both shallow and deep earthquakes, have the best adjust by a power law function for an intermediate range of values and that distribution can be rescaled in a way to explicit the power law behavior.
%In spite of our finds are similar with the results in \cite{paczuski2005spatial} we have showed the data are better fitted by a $q$-exponential function, which have a very good adjustment into data points since from the very beginning and not only for the intermediate values.

In this work we have considered the time window model presented in \cite{Ferreira2018} to construct a complex network of epicenters for deep events. 
Similarly to what was done for shallow earthquakes, our results show that the epicenter network built using the time window model proves to be better and more effective than the successive model in finding long-range relationships between the network vertices.

For deep earthquakes the time window model is also able to identify the regions with intense activity of deep earthquakes, even if these regions are not those with the largest earthquakes. This is a relevant result, given that for deep events, the occurrence of large events is not directly related to high aftershocks rates, where it can be demonstrated by the Tonga subduction zone example. 
Although the Tonga subduction zone has the largest number of deep earthquakes, South America hosts the largest-magnitude deep ruptures. 
However, while Tonga has about ten times more events with $m = 5.5$, the two regions have similar numbers of  $m = 6.5$ events, and the biggest deep earthquakes occur in South America \cite{houston2015deep}.

The present study also strengthens the idea of long-range correlations between earthquakes, once we have the presence of nonextensive characteristics in the epicenter network of deep earthquakes, as well as the scale-free and small-world properties. 
Besides that the connectivity distribution exhibit scaling properties with respect to the time  window size chosen, since that distribution can be made independent from the size of the time window, using a scale function.

% \red{An important highlight to be made here is that the results obtained so far are interesting because contributes to the.}

% \red{THOSE RESULTS MAKE SENSE ONCE LONG-RANGE.....}

These results are in close agreement with previous works that have been indicating spatial and temporal long-range correlations  in both local and worldwide views.
The outcomes of this paper also become interesting because they reinforce the criticality idea in the seismological system.

For future works, we plan to do a deeper analysis about the communities structure present in the epicenter network, as well as the implications of perform studies on spatial and temporal probabilities distributions under the NESM view. 
The results will be published elsewhere.
%communities \\
%assortative mixing \\
%hypocenter (3d)
%the present work may have continuation  of the present work is

%Comparative results showing TW model is a better approach than successive.

\section*{Acknowledgements}
J.R. thanks for the scholarship of FAPERJ, Brazil in this research, P.S.L.O thanks for the scholarship of CNPq, Brazil, R.P.F thanks FAPERJ, Brazil for productivity grant, A.R.R.P thanks CNPq, Brazil for productivity grant and D.S.R.F, R.P.F and A.R.P thank FAPERJ, Brazil for grant number 241029.

%%%%%%%%%%%%%%%%%%%%%%%
%% Elsevier bibliography styles
%%%%%%%%%%%%%%%%%%%%%%%
%% To change the style, put a % in front of the second line of the current style and
%% remove the % from the second line of the style you would like to use.
%%%%%%%%%%%%%%%%%%%%%%%

%% natbib.sty is loaded by default. However, natbib options can be
%% provided with \biboptions{...} command. Following options are
%% valid:

%%   round  -  round parentheses are used (default)
%%   square -  square brackets are used   [option]
%%   curly  -  curly braces are used      {option}
%%   angle  -  angle brackets are used    <option>
%%   semicolon  -  multiple citations separated by semi-colon
%%   colon  - same as semicolon, an earlier confusion
%%   comma  -  separated by comma
%%   numbers-  selects numerical citations
%%   super  -  numerical citations as superscripts
%%   sort   -  sorts multiple citations according to order in ref. list
%%   sort&compress   -  like sort, but also compresses numerical citations
%%   compress - compresses without sorting
%%
%% \biboptions{comma,round}

% \biboptions{}

%% Numbered

%% Numbered without titles
%\bibliographystyle{model1a-num-names}

%% Harvard
%\bibliographystyle{model2-names.bst}\biboptions{authoryear}

%% Vancouver numbered
%\usepackage{numcompress}\bibliographystyle{model3-num-names}

%% Vancouver name/year
%\usepackage{numcompress}\bibliographystyle{model4-names}\biboptions{authoryear}

%% APA style
%\bibliographystyle{model5-names}\biboptions{authoryear}

%% AMA style
%\usepackage{numcompress}\bibliographystyle{model6-num-names}

\section*{References}
\bibliographystyle{elsarticle-num} %% `Elsevier LaTeX' style
\bibliography{TW_deep_bibfile}

\begin{thebibliography}{10}
\expandafter\ifx\csname url\endcsname\relax
  \def\url#1{\texttt{#1}}\fi
\expandafter\ifx\csname urlprefix\endcsname\relax\def\urlprefix{URL }\fi
\expandafter\ifx\csname href\endcsname\relax
  \def\href#1#2{#2} \def\path#1{#1}\fi

\bibitem{gutenberg1942}
B.~Gutenberg, C.~Richter, Earthquake magnitude, intensity, energy, and
  acceleration, Bull. Seism. Soc. Am. 32~(3) (1942) 163--191.

\bibitem{omori1894aftershocks}
F.~Omori, J.~Coll, On the aftershocks of earthquakes, Sci. Imp. Univ. Tokyo 7
  (1894) 111--200.

\bibitem{foster2001competitive}
J.~Foster, Competitive selection, self-organisation and joseph a. schumpeter,
  in: Capitalism and Democracy in the 21st Century, Springer, 2001, pp.
  317--334.

\bibitem{hidalgo2009building}
C.~A. Hidalgo, R.~Hausmann, The building blocks of economic complexity,
  Proceedings of the National Academy of Sciences 106~(26) (2009) 10570--10575.

\bibitem{adamic2000power}
L.~A. Adamic, B.~A. Huberman, Power-law distribution of the world wide web,
  Science 287~(5461) (2000) 2115--2115.

\bibitem{pukdeboon2016anti}
C.~Pukdeboon, Anti-disturbance inverse optimal control for spacecraft position
  and attitude maneuvers with input saturation, Advances in Mechanical
  Engineering 8~(5) (2016) 1687814016649887.

\bibitem{anchang2009modeling}
B.~Anchang, M.~J. Sadeh, J.~Jacob, A.~Tresch, M.~O. Vlad, P.~J. Oefner,
  R.~Spang, Modeling the temporal interplay of molecular signaling and gene
  expression by using dynamic nested effects models, Proceedings of the
  National Academy of Sciences 106~(16) (2009) 6447--6452.

\bibitem{baianu2007categorical}
I.~C. Baianu, R.~Brown, J.~Glazebrook, Categorical ontology of complex
  spacetime structures: the emergence of life and human consciousness,
  Axiomathes 17~(3-4) (2007) 223--352.

\bibitem{vlad2009kinetic}
M.~O. Vlad, A.~D. Corlan, F.~Mor{\'a}n, R.~Spang, P.~Oefner, J.~Ross, Kinetic
  laws, phase--phase expansions, renormalization group, and inr calibration,
  Proceedings of the National Academy of Sciences 106~(16) (2009) 6465--6470.

\bibitem{abe2004scalefree}
S.~Abe, N.~Suzuki, Scale-free network of earthquakes, Europhys. Lett. 65~(4)
  (2004) 581--586.

\bibitem{abe2004small}
S.~Abe, N.~Suzuki, Small-world structure of earthquake network, Physica A
  337~(1) (2004) 357--362.

\bibitem{Abe2006}
S.~Abe, N.~Suzuki, {Complex-network description of seismicity}, Nonlinear
  Processes in Geophysics 13~(2) (2006) 145--150.

\bibitem{Abe2007}
S.~Abe, N.~Suzuki, {Dynamical evolution of clustering in complex network of
  earthquakes}, European Physical Journal B 59~(1) (2007) 93--97.

\bibitem{Abe2011}
S.~Abe, D.~Past{\'{e}}n, N.~Suzuki, {Finite data-size scaling of clustering in
  earthquake networks}, Physica A: Statistical Mechanics and its Applications
  390~(7) (2011) 1343--1349.

\bibitem{Lotfi2012}
N.~Lotfi, A.~H. Darooneh, {The earthquakes network: The role of cell size},
  European Physical Journal B 85~(1) (2012) 10--13.

\bibitem{Pasten2018}
D.~Past{\'{e}}n, F.~Torres, B.~A. Toledo, V.~Mu{\~{n}}oz, J.~Rogan, J.~A.
  Valdivia, {Non-universal critical exponents in earthquake complex networks},
  Physica A: Statistical Mechanics and its Applications 491 (2018) 445--452.

\bibitem{Chorozoglou2019}
D.~Chorozoglou, E.~Papadimitriou, D.~Kugiumtzis, {Investigating small-world and
  scale-free structure of earthquake networks in Greece}, Chaos, Solitons and
  Fractals 122 (2019) 143--152.

\bibitem{he2019statistical}
X.~He, L.~Wang, H.~Zhu, Z.~Liu, Statistical properties of complex network for
  seismicity using depth-incorporated influence radius, Acta Geophysica (2019)
  1--9.

\bibitem{Peixoto2004a}
T.~P. Peixoto, C.~P. Prado, {Distribution of epicenters in the
  Olami-Feder-Christensen model}, Physical Review E - Statistical, Nonlinear,
  and Soft Matter Physics 69~(2 2) (2004) 1--4.

\bibitem{Peixoto2004b}
T.~P. Peixoto, C.~P. Prado, {Statistics of epicenters in the
  Olami-Feder-Christensen model in two and three dimensions}, Physica A:
  Statistical Mechanics and its Applications 342~(1-2 SPEC. ISS.) (2004)
  171--177.

\bibitem{Peixoto2006}
T.~P. Peixoto, C.~P.~C. Prado, {Network of epicenters of the
  Olami-Feder-Christensen model of earthquakes}, Physical Review E -
  Statistical, Nonlinear, and Soft Matter Physics 74~(1) (2006) 1--9.

\bibitem{Caruso2006}
F.~Caruso, V.~Latora, A.~Pluchino, A.~Rapisarda, B.~Tadi{\'{c}},
  {Olami-Feder-Christensen model on different networks}, European Physical
  Journal B 50~(1-2) (2006) 243--247.

\bibitem{ferreira2015agreement}
D.~S. Ferreira, A.~R. Papa, R.~Menezes, On the agreement between
  small-world-like {OFC} model and real earthquakes, Physics Letters A 379~(7)
  (2015) 669--675.

\bibitem{steeples1996far}
D.~W. Steeples, D.~D. Steeples, Far-field aftershocks of the 1906 earthquake,
  Bulletin of the Seismological Society of America 86~(4) (1996) 921--924.

\bibitem{Baiesi2004}
M.~Baiesi, M.~Paczuski, {Scale-free networks of earthquakes and aftershocks},
  Physical Review E 69~(6) (2004) 8.

\bibitem{baiesi2005complex}
M.~Baiesi, M.~Paczuski, Complex networks of earthquakes and aftershocks,
  Nonlinear Processes in Geophysics 13 (2005) 1--11.

\bibitem{paczuski2005spatial}
J.~Davidsen, M.~Paczuski, Analysis of the spatial distribution between
  successive earthquakes, Physical Review Letters 94~(048501) (2005)
  3747--3752.

\bibitem{Abe2012}
S.~Abe, N.~Suzuki, {Universal law for waiting internal time in seismicity and
  its implication to earthquake network}, EPL 97~(4) (2012).

\bibitem{Bendick2017}
R.~Bendick, R.~Bilham, {Do weak global stresses synchronize earthquakes?},
  Geophysical Research Letters 44~(16) (2017) 8320--8327.

\bibitem{ferreira2014small}
D.~S. Ferreira, A.~R. Papa, R.~Menezes, Small world picture of worldwide
  seismic events, Physica A: Statistical Mechanics and its Applications 408
  (2014) 170--180.

\bibitem{Ferreira2018}
D.~Ferreira, J.~Ribeiro, A.~Papa, R.~Menezes, Towards evidence of long-range
  correlations in shallow seismic activities, EPL 121~(5) (2018) 58003.

\bibitem{frohlich2006deep}
C.~Frohlich, Deep earthquakes, Cambridge University Press, 2006.

\bibitem{blondel2008fast}
V.~D. Blondel, J.-L. Guillaume, R.~Lambiotte, E.~Lefebvre, Fast unfolding of
  communities in large networks, J. Stat. Mech. Theor. Exp. 10 (2008) 1000.

\bibitem{tsallis1988possible}
C.~Tsallis, Possible generalization of boltzmann-gibbs statistics, Journal of
  Statistical Physics 52~(1-2) (1988) 479--487.

\bibitem{barrat2004architecture}
A.~Barrat, M.~Barthelemy, R.~Pastor-Satorras, A.~Vespignani, The architecture
  of complex weighted networks, Proceedings of the National Academy of Sciences
  of the United States of America 101~(11) (2004) 3747--3752.

\bibitem{brandes2001faster}
U.~Brandes, A faster algorithm for betweenness centrality, Journal of
  Mathematical Sociology 25~(2) (2001) 163--177.

\bibitem{barabasi2016network}
A.-L. Barab{\'a}si, et~al., Network science, Cambridge university press, 2016.

\bibitem{wiens2001seismological}
D.~A. Wiens, Seismological constraints on the mechanism of deep earthquakes:
  Temperature dependence of deep earthquake source properties, Physics of the
  Earth and Planetary Interiors 127~(1-4) (2001) 145--163.

\bibitem{houston2015deep}
H.~Houston, Deep earthquakes, Treatise on Geophysics 4 (2015) 329--354.

\end{thebibliography}

\end{document}